\documentstyle[preprint,aps,prd]{revtex}
\begin{document}
\preprint{MRI-PHY/P20000309}
\title{Nonstatic global string in Brans-Dicke theory}
\author{A.A.Sen$^{1}$\footnote{anjan@mri.ernet.in} and 
N.Banerjee$^{2}$\footnote{narayan@juphys.ernet.in}}
\address{$^{1}$Mehta Research Institute,Chhatnag Road,Jhushi\\
Allahabad 211019,India\\
$^{2}$Relativity and Cosmology Research Centre,\\
Department of Physics, Jadavpur University,\\
Calcutta 700032, India}
\maketitle
\begin{abstract}
Gravitational field of a nonstatic global string has been studied in the context 
of Brans-Dicke theory of gravity. Both the metric components and the BD scalar field are
assumed to be nonseparable functions of time and space.The spacetime may or may not have
any singularity at a finite distance from the string core but the singularity 
at a particular time always remains. It has been shown that the spacetime exhibits both
outgoing and incoming gravitational radiation.
\end{abstract}
\section{Introduction}

In the recent years there has been a rapid growth of research in the interface between
particle physics and cosmology resulting in several exciting ideas concerning the
unification of forces of nature. One of the most important features in the early
universe, particle physicists and cosmologists have speculated, is the "Grand
Unification"- unification of the strong and electroweak forces. It was soon realised
that this grand symmetery presented in the early universe at very high energy scale
had to be broken in  order to account for the electroweak and nuclear forces as
different fields in the present day low energy universe. One of the immediate and
inescapable consequences of this symmetry breaking is the formation of topological
defects. They can be monopoles, cosmic strings or domain walls~\cite{Kibble}.

Among these defects, cosmic strings are particularly interesting in view of their
capability to produce direct observational effects such as gravitational lensing and
also because they are possible seeds of galaxy formation~\cite{VS}. Strings are said
to be local or global depending on their origin from the breakdown of local or global
$U(1)$ symmetry. While local strings are well behaved~\cite{local}, having an exterior
representing a flat Minkowskian spacetime with a conical defect, global strings,
however, have strong gravitational effects at large distances. The static global
string spaectime is found to have a singularity at a finite distance~\cite{global}
from the string core. Later Gregory~\cite{Greg1} showed that the time dependence of
the metric might remove  the singular behaviour of the global string spacetime.Very
recently Sen and Banerjee have shown that one can also have nonsingular static
spacetime outside the core of the global string if the spacetime does not admit
Lorentz boost along the symmetry axis~\cite{Sen1}.

Discussions in the previous paragraphs have been confined within the context of
General Relativity(GR). But at sufficiently high energy scales, it seems likely that
gravity is not given by the Einstein action but becomes modified by the superstring
terms. In the low energy limits of this string theory, one recovers Einstein's gravity
along with scalar dilaton field which is nonminimally coupled to the
gravity~\cite{GSW}. On the other hand, scalar tensor theories, such as Brans-Dicke
theory(BD)~\cite{BD}, which is compatible with the Mach's principle, have been
considerably
revived in the recent years. In these theories, the purely metric coupling of matter
with gravity is restored and hence the equivalence principle is ensured. Moreover these
models exhibit an attractor mechanism towards GR, so that the expansion of the
universe during matter dominated epoch tends to drive the scalar field towards a state
where these models are indistinguishable from GR~\cite{DN}. Although dilaton gravity
and BD theory arise from entirely different motivations, it can be shown that the
former is special case of the later, at least formally~\cite{Sen2}. The implications
of the BD theory for topological defects have been studied by many authors. Gundlach
and Ortiz~\cite{GO} obtained analytical solutions for a local gauge string in BD
theory. These solutions are not satisfactory because here the BD theory was used to
solve for the exterior, while in the interior, the solutions were given for Einstein's
field equations, ignoring the BD scalar field completely. The other solutions in BD
theory for a local gauge string have been given by Barros and Romero~\cite{BR} and
Guimaraes~\cite{GM}. Later Sen et.al~\cite{Sen3} have shown that for a local gauge
string, Vilenkin's prescription for the energy momentum tensor due to the string is
inconsistent with BD theory. Gravitational field of a $U(1)$ global string has also
been studied by many authors. Sen et.al~\cite{Sen3} have presented two kind of
solutions for the specetime outside the core of a static global string, one of which
was in closed form and was nonsingular  at a finite distance from the string core
while the other one could not be written in a closed form and moreover contained
singularity at a finite distance from the string core. More general solutions for
static global string in BD theory have been provided by Boisseau and Linet~\cite{BL}. In
another interesting work, Dando and Gregory~\cite{DG} have examined field equations of
a self gravitating nonstatic global string in low energy string gravity allowing for an
arbitary coupling of the global string to the dilaton field. Both massive and massless
dilaton fields were considered. In both cases, they had demonstrated the existence of
a nonsingular spacetime for the string if one includes a time dependence of the form
$e^{b_{0}t}$ along the length of the string where $b_{0}>0$. But here the dilaton
scalar field itself was not time dependent. In this work, we have studied the
gravitational field of a nonstatic global string in BD theory where the BD scalar
field is also time depenedent along with the metric components. Moreover the metric 
components are not separable functions of time and space which was the assumption for
most of the previous works.

In section II, we have constructed the field equations for the time dependent global string in BD
theory and have presented the solutions for the field equations. In section III, we have analysed
our solutions where we have studied the presence of the physical singularity in our spacetime and
also have shown that the spacetime represents both incoming and outgoing gravitational radiations.
The paper ends with a conclusion in section IV.

 \section{Field equations and their solutions}

The gravitational field equation in BD theory is given by ~\cite{BD}
$$
G_{\mu\nu}={T_{\mu\nu}\over{\Phi}}+{\omega\over{\Phi^{2}}}(\Phi_{,\mu}\Phi_{,\nu}-
{1\over{2}}g_{\mu\nu}\Phi_{,\alpha}\Phi^{,\alpha})+
{1\over{\Phi}}(\Phi_{,\mu;\nu}-g_{\mu\nu}\Box\Phi)
\eqno{(1.1)},
$$
together with the wave equation for the BD scalar field $\Phi$
$$
\Box\Phi={T\over{(2\omega+3)}}
\eqno{(1.2)},
$$
where $\Phi$ is the BD scalar field, $T_{\mu\nu}$ is the energy momentum tensor for 
the matter field and $\omega$ is the BD parameter. The conservation relation for the 
matter field $T^{\mu\nu}_{;\nu}=0$ follows identically. Sometimes it is more convenient 
to use a nonphysical metric ${\bar{g}}_{\mu\nu}$ which is conformally related to 
$g_{\mu\nu}$ by ~\cite{DIC}
$$
g_{\mu\nu} = \Phi^{-1}{\bar{g}}_{\mu\nu}
\eqno{(1.3)},
$$
and the corresponding  field equations are 
$$
{\bar{G}}_{\mu\nu} = {\bar{T}}_{\mu\nu} +
{(2\omega+3)\over{2\Phi^{2}}}(\Phi_{,\mu}\Phi_{,\nu}-
{1\over{2}}{\bar{g}}_{\mu\nu}\Phi^{,\alpha}\Phi_{,\alpha})
\eqno{(1.4)},
$$
$$
{{\Box}}\Phi = {{\bar{T}}\over{2\omega+3}}
\eqno{(1.5)},
$$
where $\Box$ operator is calculated with respect to ${\bar{g}}_{\mu\nu}$ and also now 
${\bar{T}}_{\mu\nu}$ is the conformally transformed energy momentum tensor
given by
$$
T_{\mu\nu} = \Phi{\bar{T}}_{\mu\nu}
\eqno{(1.6)}.
$$
But now the conservation equation for the matter field does not follow identically.
The rest mass of the particles vary depending on the spacetime locations
and hence the paths of the massive particles are no longer geodesics.

We have taken the line element as 
$$
ds^{2} = e^{2(k-u)}(dt^{2}-dr^{2})-e^{2u}dz^{2}-w^{2}e^{-2u}d\theta^{2}
\eqno{(1.7)},
$$
where $k,u,w$ are functions of both $t$ and $r$. Assuming that the complex
scalar field for the global string is independent of the radial distance
$r$ outside the core radius $\delta=(\eta\sqrt{\lambda})^{-1}$, where
$\eta$ is the symmetry breaking energy scale and $\lambda$ is a constant,
one can calculate the energy momentum tensor for the global string 
in the original version of BD theory as
$$
T^{t}_{t}=T^{r}_{r}=T^{z}_{z}=-T^{\theta}_{\theta}={\eta^{2}\over{2}}{e^{2u}
\over{w^{2}}}
\eqno{(1.8)}.
$$
In our case the $T^{\mu}_{\nu}$s are conformally transformed as (1.6),
hence we assume ${\bar{T}}^{\mu}_{\nu}$ in the conformally transformed
version as 
$$
{\bar{T}}^{t}_{t}={\bar{T}}^{r}_{r}={\bar{T}}^{z}_{z}=-{\bar{T}}^{\theta}_
{\theta}={\bar{\sigma}}(r,t)
\eqno{(1.9)}
$$
With (1.7) and (1.9), equations (1.4) and (1.5) become
$$
{\dot{k}\dot{w}\over{w}}-{\dot{u}}^{2}-{w^{''}\over{w}}+{k^{'}w^{'}\over{w}}
-u^{'2}=-{\bar{\sigma}}e^{2(k-u)}-{(2\omega+3)\over{4}}(\dot{\psi}^{2}+
\psi^{'2})
\eqno{(1.91a)}
$$
$$
{\ddot{w}\over{w}}-{\dot{k}\dot{w}\over{w}}+{\dot{u}}^{2}-{k^{'}w^{'}\over{w}}
+u^{'2}=-{\bar{\sigma}}e^{2(k-u)}+{(2\omega+3)\over{4}}(\dot{\psi}^{2}+
\psi^{'2})
\eqno{(1.91b)}
$$
$$
{\dot{u}}^{2}+{\ddot{k}}-u^{'2}-k^{''}={\bar{\sigma}}e^{2(k-u)}-
{(2\omega+3)\over{4}}(\dot{\psi}^{2}-
\psi^{'2})
\eqno{(1.91c)}
$$
$$
{\ddot{w}\over{w}}-{2\dot{u}\dot{w}\over{w}}-2{\ddot{u}}+{\dot{u}}^{2}+
{\ddot{k}}-{w^{''}\over{w}}+{2u^{'}w^{'}\over{w}}+2u^{''}-k^{''}-u^{'2}=
-{\bar{\sigma}}e^{2(k-u)}+{(2\omega+3)\over{4}}(\dot{\psi}^{2}-
\psi^{'2})
\eqno{(1.91d)}
$$
$$
{k^{'}{\dot{w}}\over{w}}-{\dot{w}^{'}\over{w}}+{\dot{k}w^{'}\over{w}}-
2u^{'}\dot{u}=-{(2\omega+3)\over{2}}\dot{\psi}\psi^{'}
\eqno{(1.91e)}
$$
$$
\ddot{\psi}+{\dot{\psi}\dot{w}\over{w}}-\psi^{'}-{\psi^{'}w^{'}\over{w}}
={2{\bar{\sigma}}e^{2(k-u)}\over{(2\omega+3)}}
\eqno{(1.91f)}
$$
where we have assumed $\psi=ln(\Phi)$ and overdot and prime represent differentiation
with respect to $t$ and $r$ respectively.
After some straightforward calculations one can get
$$
{(\dot{u}w)}^{.}=(u^{'}w)^{'}
\eqno{(1.92)}.
$$
One of the possible solutions of (1.92) is that both $(\dot{u}w)$ and $(u^{'}w)$ are
functions of $(r+t)$. In what follows, we have assumed that $k, u, w, \psi, \sigma$ are 
functions of $(at+br)$ in our subsequent calculations where $a, b$ are arbitary
constants.

Defining $x=at+br$, equations (1.91a)-(1.91f) become
$$
(a^{2}+b^{2}){k_{x}w_{x}\over{w}}-(a^{2}+b^{2})u_{x}^{2}-b^{2}{w_{xx}\over{w}}=
-{\bar{\sigma}}e^{2(k-u)}-{(2\omega+3)\over{4}}(a^{2}+b^{2})\psi_{x}^{2}
\eqno{(1.93a)}
$$
$$
a^{2}{w_{xx}\over{w}}-(a^{2}+b^{2}){k_{x}w_{x}\over{w}}+(a^{2}+b^{2})u_{x}^{2}=
-{\bar{\sigma}}e^{2(k-u)}+{(2\omega+3)\over{4}}(a^{2}+b^{2})\psi_{x}^{2}
\eqno{(1.93b)}
$$
$$
(a^{2}-b^{2})u_{x}^{2}+(a^{2}-b^{2})k_{xx}={\bar{\sigma}}e^{2(k-u)}+
{(2\omega+3)\over{4}}(a^{2}-b^{2})\psi_{x}^{2}
\eqno{(1.93c)}
$$
$$
(a^{2}-b^{2}){w_{xx}\over{w}}-2(a^{2}-b^{2}){u_{x}w_{x}\over{w}}-2(a^{2}-b^{2})u_{xx}+
(a^{2}-b^{2})u_{x}^{2}+(a^{2}-b^{2})k_{xx}
$$
$$
\hspace{15mm}
=-{\bar{\sigma}}e^{2(k-u)}+{(2\omega+3)\over{4}}(a^{2}-b^{2})\psi_{x}^{2}
\eqno{(1.93d)}
$$
$$
(a^{2}-b^{2})(\psi_{xx}+{\psi_{x}w_{x}\over{w}})={2{\bar{\sigma}}e^{2(k-u)}
\over{2\omega+3}}
\eqno{(1.93e)}
$$
It can be shown that equation (1.91e) is now no longer indepenedent but can be achieved
with the help of other equations. After some straightforward calculations one can get
from (1.93a)-(1.93e) the following equations:
$$
(a^{2}-b^{2}){w_{xx}\over{w}}=-2{\bar{\sigma}}e^{2(k-u)}
\eqno{(1.94a)}
$$
$$
{w_{xx}\over{w}}-{2k_{x}w_{x}\over{w}}+2u_{x}^{2}={(2\omega+3)\over{2}}\psi_{x}^{2}
\eqno{(1.94b)}
$$
$$
u_{x}=\alpha/w
\eqno{(1.94c)}
$$
$$
k_{x}=\beta/w
\eqno{(1.94d)}
$$
$$
e^{\psi}=Bw^{-1/(2\omega+3)}
\eqno{(1.94e)}
$$
where $\alpha$, $\beta$ and $B$ are arbitary integration constants. So we have five
equations and five unknowns namely, $w, k, u, \psi$ and ${\bar{\sigma}}$. Also one
should note from equation (1.94a) that $a^{2}\neq b^{2}$ in order to have nonvanishing 
$\bar{\sigma}$.
Using (1.94c)-(1.94e), (1.94b) becomes
$$
ww_{xx}-2\beta w_{x}+2\alpha^{2}-{1\over{2(2\omega+3)}}w_{x}^{2}=0
\eqno{(1.95)}.
$$
To have an exact analytical solution for the equation (1.95), we make a
simplified assumption that $\alpha=\beta=0$ which means
$$
u_{x}=k_{x}=0
\eqno{(1.96)}.
$$
Then one can integrate equation (1.95) to get
$$
w=w_{0}(x-x_{0})^{1/(1-A)}
\eqno{(1.97)}
$$
where $w_{0}$ and $x_{0}$ are arbitary constants and $A={1\over{2(2\omega+3)}}$. One
can now calculate the expressions for the energy density ${\bar{\sigma}}$ and the BD
scalar field $\Phi$:
$$
\bar{\sigma}={p(p-1)(b^{2}-a^{2})\over{2(x-x_{0})^{2}}}
\eqno{(1.98a)}
$$
$$
\Phi=e^{\psi}=\Phi_{0}(x-x_{0})^{2A/(A-1)}
\eqno{(1.98b)}
$$
where we have taken the constants $e^{k}=e^{u}=1$ without any loss of generality and
$p=1/(1-A)$. Using (1.3) and (1.6) one
can now calculate the expression for the line element and the energy density in the
original BD theory which are given by
$$
ds^{2}={1\over{\Phi_{0}}}(at+br-x_{0})^{2A/(1-A)}(dt^{2}-dr^{2}-dz^{2})-
{w_{0}^{2}\over{\Phi_{0}}}(at+br-x_{0})^{2(1+A)/(1-A)}d\theta^{2}
\eqno{(1.99a)}
$$
$$
\sigma={p(p-1)(b^{2}-a^{2})\Phi_{0}^{2}\over{2}}(at+br-x_{0})^{-2(1+A)/(1-A)}
\eqno{(1.99b)}.
$$
One can check that the energy density $\sigma\propto 1/g_{33}$ which is true for 
global string according to (1.8). To recover the corresponding solution for Einstein's General
Relaivity one has to take the limit $\omega\rightarrow\infty$ \cite{BD}. In the
present case, in this limit,
 the BD scalar field $\Phi$ becomes constant and the line element (1.99a) becomes a flat metric with an angular deficit.

One can also calculate the angular deficit in the spacetime for a constant time which becomes
$$
2\pi\left [1-{bw_{0}(at+br-x_{0})^{(1+A)/(1-A)}\over{(1-A)[(at+br-x_{0})^{1/(1-A)}-
(at-x_{0})^{1/(1-A)}]}}\right].
$$
Hence it is clear from the above expression that the angular deficit is a
function of both $r$ and $t$.
\section{Analysis of the solution}

To check the occurance of the curvature singularity in the spacetime, one has to
calculate the
Kretschmann curvature scalar for the line element (1.99a):
$$
øR^{ijkl}R_{ijkl}=4\Phi_{0}^{2}\left[{(a^{2}-b^{2})^{2}(80\omega^{2}+248\omega+198)(at+br
-x_{0})^{-8(2\omega+3)/(4\omega+7)}\over{(16\omega^{2}+40\omega+25)^{2}}}\right]
\eqno{(2.1)}
$$
One can see that for $x_{0}\leq 0$, the spacetime does not have any curvature singularity at
any finite distance from the string core if $(2\omega+3)>0$ which is a physical 
assumption so as to make the energy contribution from the scalar field positive~\cite{DN}. 
There may be singularity at $r=0$ but line element (1.99a) is valid for $r>\delta$
because the form of the energy momentum tensor given in (1.8) is valid only for outside the 
string core that is $r>\delta$. 
For $x_{0}>0$ the spacetime may have singularity at a finite distance from the string core. 
But there is always a physical singularity at a particular time 
in the spacetime depending on the choice of the constant as the range of $t$  is $-\infty\leq t\leq\infty$.

To check whether the spacetime represents the emission and absorption of gravitational
radiation, one has to calculate the corresponding Weyl tensor, which is thought of as
representing the pure gravitational field~\cite{SW}. Introducing the null tetrad frame by
$$
L^{\mu}={e^{u-k}\over{\sqrt{2}}}(\delta^{\mu}_{t}-\delta^{\mu}_{r})
\eqno{(2.2a)}
$$
$$
N^{\mu}={e^{u-k}\over{\sqrt{2}}}(\delta^{\mu}_{t}+\delta^{\mu}_{r})
\eqno{(2.2b)}
$$
$$
M^{\mu}={1\over{\sqrt{2}}}(e^{-u}\delta^{\mu}_{z}+{i\over{w}}\delta^{\mu}_{\theta})
\eqno{(2.2c)}
$$
$$
{\bar{M}}^{\mu}={1\over{\sqrt{2}}}(e^{-u}\delta^{\mu}_{z}-{i\over{w}}\delta^{\mu}_{\theta})
\eqno{(2.2d)}
$$
one can find that the two of the nonvanishing components of the Weyl tensor for
the line element
(1.99a)
are
$$
\Psi_{0}=-C_{\mu\nu\lambda\sigma}L^{\mu}M^{\nu}L^{\lambda}M^{\sigma}
\eqno{(2.3a)},
$$
$$
\Psi_{4}=-C_{\mu\nu\lambda\sigma}N^{\mu}{\bar{M}}^{\nu}N^{\lambda}{\bar{M}}^{\sigma}
\eqno{(2.3b)}.
$$
The reason for calculating these two components in null frame is that these components in
this frame have a direct physical meaning that is $\Psi_{0}$ represents an outgoing
cylindrical gravitational wave along the null hypersurface defined by $N^{\mu}$ and
$\Psi_{4}$ represents the incoming cylindrical gravitational wave propagating along the null
hypersuface
defined by $L^{\mu}$~\cite{KSM}. Using the line element (1.99a) and (2.2)
and
(2.3) these components become
$$
\Psi_{0}={3A\Phi_{0}^{3}(a^{2}+b^{2})(at+br-x_{0})^{-2(1+2A)(1-A)}\over{24(1-A)^{2}}}\left[
{\Phi_{0}(at+br-x_{0})^{-2(1+A)/(1-A)}\over{w_{0}^{2}}}+1\right]
$$
$$
-{abA\Phi_{0}^{3}(a^{2}+b^{2})(at+br-x_{0})^{-2(1+2A)(1-A)}\over{24(1-A)^{2}}}\left[
{\Phi_{0}(at+br-x_{0})^{-2(1+A)/(1-A)}\over{w_{0}^{2}}}+1\right]
\eqno{(2.4a)}
$$
$$
\Psi_{4}={3A\Phi_{0}^{3}(a^{2}+b^{2})(at+br-x_{0})^{-2(1+2A)(1-A)}\over{24(1-A)^{2}}}\left[
{\Phi_{0}(at+br-x_{0})^{-2(1+A)/(1-A)}\over{w_{0}^{2}}}+1\right]
$$
$$
+{abA\Phi_{0}^{3}(a^{2}+b^{2})(at+br-x_{0})^{-2(1+2A)(1-A)}\over{24(1-A)^{2}}}\left[
{\Phi_{0}(at+br-x_{0})^{-2(1+A)/(1-A)}\over{w_{0}^{2}}}+1\right]
\eqno{(2.4b)}
$$ 
The first terms in the expressions for $\Psi_{0}$ and $\Psi_{4}$ are equal and are
nonvanishing for $a=0$ i.e for static spacetime. They represent the gravitational field for
the global string. The second terms in $\Psi_{0}$ and $\Psi_{4}$ are of opposite sign and
does not exist for $a=0$ i.e for static spacetime. These terms represent the outgoing and
incoming gravitational radiation respectively. For a particular case $ab=3$ one can see that there is no outgoing radiation from the string but only the incoming radiation exists.
\section{Conclusion}

In this work we have studied the gravitational field outside the core of a nonstatic global string
in BD theory of gravity. The solution presented here are not the general one as we have made 
some assumptions but
this may be the first example of such investigations where the BD scalar field
is itself time dependent and also where the metric components are not the separable function
of $t$ and $r$. The previous work in this line by Dando and Gregory~\cite{DG},
assumed the metric components are separable functions of time and space and also the BD
scalar field was time independent. The solution presented here may or may not have singularity at
a finite distance from the string core depending upon the value of arbitary constant but it
has singularity at a particular time. One can also check from the expression of Kretschmann scalar that the
spaceime is asymtotically flat both in time and space for $(2\omega+3)>0$ which is a physical
assumption~\cite{BD}. As the components of Weyl tensor in null frame are nonzero hence there
are
incoming as well as outgoing gravitational radiation in our spacetime. Also the fact that the weyl
tensor has non zero components combined with the non zero Ricci tensor leads to the fact that
these strings may produce some gravitational lens effects for null rays passing near
the strings. Hence there may be distortion and amplifications of distance objects seen along the
lines of sight passing near the string~\cite{DM}.
\section{Acknowledgement}
One of the authors (AAS) gratefully acknowledge the hospitality provided by the Abdus Salam
International Center for Theoretical Physics, Trieste, Italy,  where part of the
work has been done.

\end{document}